\documentclass[a4paper,11pt]{article}
\usepackage{jcappub} 

\usepackage{array} 
\usepackage{booktabs} 

\usepackage{hyperref}
\hypersetup{colorlinks=true,linkcolor=blue,citecolor=cyan}
\usepackage{graphicx}
\usepackage{xcolor, soul}
\sethlcolor{green}
\usepackage{dcolumn}
\usepackage{bm}
\usepackage{color}
\usepackage{enumitem}
\usepackage{mathrsfs}
\usepackage{amsmath}
\usepackage{amssymb}
\usepackage{cleveref}
\usepackage{orcidlink}

\crefname{equation}{Eqn.}{Eqns.}
\crefname{figure}{Fig.}{Figs.}
\crefname{section}{Sec.}{Sec.}
\crefname{table}{Table}{Tables}
\usepackage{physics}
\usepackage{multirow}
\setstcolor{red}

\arxivnumber{2412.03885} 
\title{
\textcolor{black}{Energy Extraction from Loop Quantum Black Holes: The Role of Magnetic Penrose Process and Quantum Gravity Effects with Astrophysical Insights}}

\author[a]{Tursunali Xamidov}
\affiliation[a]{Institute of Fundamental and Applied Research, National Research University TIIAME, Kori Niyoziy 39, Tashkent 100000, Uzbekistan} 

\author[b]{Pankaj Sheoran
}
\affiliation[b]{Department of Physics, School of Advanced Sciences, Vellore Institute of Technology, Tiruvalam Rd, Katpadi, Vellore, Tamil Nadu 632014, India}

\author[c,d,e,f]{Sanjar Shaymatov}
\affiliation[c]{Institute for Theoretical Physics \& Cosmology, Zhejiang University of Technology, Hangzhou 310023, China}
\affiliation[d]{University of Tashkent for Applied Sciences, Str. Gavhar 1, Tashkent 100149, Uzbekistan}
\affiliation[f]{Western Caspian University, Baku AZ1001, Azerbaijan}

\author[c,g]{Tao Zhu}
\affiliation[g]{United Center for Gravitational Wave Physics (UCGWP), Zhejiang University of Technology, Hangzhou 310023, China}

\emailAdd{xamidovtursunali@gmail.com}
\emailAdd{pankaj.sheoran@vit.ac.in}
\emailAdd{sanjar@astrin.uz}
\emailAdd{zhut05@zjut.edu.cn}

 \abstract{
 In this study, we explore the influence of quantum gravitational corrections, derived from Loop Quantum Gravity (LQG), on the efficiency of the magnetic Penrose process (MPP) in black hole (BH) environments. We begin by analyzing the rotating Loop Quantum Black Hole (LQBH) metric, describing the structure of the event horizon and ergosphere as functions of the quantum parameter $\epsilon = \gamma \delta$, with $\gamma$ representing the Immirzi parameter and $\delta$ the polymeric parameter, and the spin parameter $a$. These modifications provide a novel setting for exploring the dynamics of charged particles near the LQBH and evaluating the resultant energy extraction through the MPP. Interestingly, for a given value of the LQBH parameter $a$, we observe that the ergosphere region of the LQBH exhibits a more intricate structure compared to its classical counterpart, the Kerr BH, as $\epsilon$ increases. Furthermore, we find that the overall efficiency of the process decreases with $\epsilon$ that decreases $a_{max}$, again in contrast to the Kerr BH, where efficiency rises with an increasing $a$.
Our analysis also extends to astrophysical contexts, applying constraints on the mass and magnetic field of LQBHs for astrophysical BH candidates, including Sgr A*, M87*, NGC 1052, and BZ (Blandford and Znajek sources, i.e., supermassive BHs with masses around $10^9 M_\odot$ and magnetic fields in the range $10^3-10^4 \text{G}$). We assess these sources as potential accelerators of high-energy protons across different values of the quantum parameter $\epsilon$. Additionally, we examine how variations in the magnetic field strength $B$ and quantum corrections impact the energy of protons accelerated from M87$^{\star}$ and Sgr A$^{\star}$ following beta decay. 
Finally, the results reveal potential observational signatures of LQG and insights into quantum gravity's role in high-energy astrophysics.
}
\begin{document}
\maketitle
\flushbottom

\section{Introduction}
\label{introduction}

Black holes (BHs), among the most fascinating and enigmatic objects in theoretical physics, play a crucial role in both theoretical and observational studies. Extensively explored through both classical and quantum frameworks, BHs provide insights into the fundamental nature of spacetime. General Relativity (GR), which describes gravity as the curvature of spacetime, offers a robust framework for understanding the macroscopic structure of BHs. Specifically, GR successfully explains features like the event horizon, gravitational waves, and BH thermodynamics. However, despite GR's success in describing the large-scale behavior of BHs, it falls short of addressing the quantum mechanical aspects of spacetime at microscopic scales, especially near singularities and the Planck scale \cite{Ashtekar:2021kfp}. This limitation underscores the need for a unified theory that reconciles quantum mechanics with gravity, highlighting the importance of BHs as a bridge between these foundational aspects of modern physics.

Loop Quantum Gravity (LQG) \cite{Ashtekar:2004eh,Rovelli:2004tv,Ashtekar:2021kfp,Rovelli:2008zza} is a leading approach to quantum gravity, which postulates that spacetime is discrete at the Planck scale, formed of quantized loops of gravitational fields. LQG modifies the classical description of BHs by introducing quantum corrections to their structure \cite{Gambini:2008dy,Perez:2017cmj,Gambini:2013hna}. These quantum corrections affect the geometry near the event horizon and the singularity, offering the possibility of a ``regularized" BH \cite{Fernandes:2020rpa,Hu:2023iuw,Lan:2023cvz} that avoids the classical singularity problem. 

One such effective quantum BH, which is derived from the mini-superspace approach in quantizing the Schwarzschild BH spacetime using the polymerization procedure in LQG, was explored in Ref.~\cite{Modesto:2008im}. This loop quantum black hole (LQBH) has no curvature singularity and its regularity is determined by two parameters arising from LQG: the polymeric parameter ($\delta$) and the Barbero-Immirzi parameter $\gamma$ \cite{Modesto:2009ve, Sahu:2015dea}. This LQBH is also known as a self-dual BH since it exhibits a T-duality symmetry (for details see, e.g., Refs.~\cite{Modesto:2009ve, Sahu:2015dea}). The introduction of parameters such as the Immirzi parameter ($\gamma$) and the polymeric parameter ($\delta$) within LQG \cite{Konoplya_2016,Ni_2016} offers a novel framework for understanding the dynamical behavior of BHs, including their mass, spin, and magnetic properties. The corresponding rotating LQBH was obtained in Ref. \cite{Liu:2020ola} using the modified Newman-Janis procedure. It is interesting to explore the quantum effects of LQBH in a wide range of astrophysical environments, see Refs. \cite{Uktamov:2024ckf, Sahu:2015dea, Zhu:2020tcf, Liu:2020ola, Tu:2023xab, Liu:2023vfh, Jiang:2023img, Jiang:2024vgn, Jiang:2024cpe, Yan:2022fkr, Yan:2023vdg, Momennia:2022tug} and references therein.

One of the most intriguing phenomena surrounding BHs is the Penrose process \cite{Penrose:1969pc,Penrose1971}, which was first proposed by Roger Penrose in 1971 as a mechanism for extracting energy from BHs. In the classical version of the process, a particle near the event horizon of a rotating BH splits into two parts: one falls into the BH, while the other escapes. The particle falling into the BH carries negative energy, which reduces the BH's spin, while the particle escaping the BH can carry away positive energy, resulting in net energy extraction \cite{Bhat1985}. This process is thought to occur in the ergosphere, a region outside the event horizon where the BH’s rotation drags spacetime itself, facilitating such particle interactions \cite{Prabhu10,Abdujabbarov11,Nozawa05}. On the other hand, the magnetic Penrose process (MPP) \cite{Wagh1985ApJ,Wagh:1989zqa,Koide:2003fj,Shaymatov24MPP1,Shaymatov24MPP2}, a variant of this mechanism, occurs in the presence of a magnetic field and plays a crucial role in the energy dynamics of astrophysical BHs. It is particularly relevant in the context of BH accretion disks and relativistic jets observed in Active Galactic Nuclei (AGN), where strong magnetic fields are believed to assist in the extraction of energy from the BH \cite{McKinney:2012wd}. The efficiency of the MPP, which is influenced by factors such as the BH’s spin and the magnetic field configuration, is thought to be responsible for some of the most energetic phenomena observed in the universe, such as high-energy particle acceleration \cite{Banados:2009pr} and relativistic jets \cite{Blandford:2018iot}.


However, much of the current understanding of the MPP relies on a classical description of BHs. The role of quantum effects, specifically those predicted by gravity theories motivated by quantum mechanics such as LQG, in modifying these energy extraction mechanisms remains largely unexplored. Given that LQG proposes a fundamentally different description of BHs, particularly in the region near the event horizon, it is essential and interesting to investigate how quantum gravitational effects might alter the MPP. In this context, the Immirzi and polymeric parameters in LQG could potentially influence the geometry of the BH, modifying both the ergosphere and the event horizon, and thereby affecting the efficiency of energy extraction. For instance, quantum corrections to the Schwarzschild geometry can significantly impact geodesic motion and energy extraction processes \cite{Battista:2023iyu}. Similarly, the spinning LQG BH model, which is singularity-free and allows for spins greater than those of Kerr BHs, shows promise as a cosmic particle accelerator, further emphasizing the influence of quantum gravity parameters on particle dynamics and energy extraction mechanisms \cite{Suresh:2024hth}.

The aim of this paper is to explore the role of LQG in the MPP, with a particular focus on how quantum corrections to the BH geometry impact the efficiency of energy extraction. 
Additionally, the astrophysical relevance of this study is significant, as it could provide new insights into the energy dynamics of astrophysical BHs, such as those found at the centres of galaxies. By constraining the mass and magnetic field of LQGBHs for well-known BH candidates such as Sgr A* \cite{EventHorizonTelescope:2022wkp,Eatough_2013}, M87* \cite{MF:2021ApJ}, NGC 1052 \cite{Baczko16}, and BZ \cite{Blandford1977}, we examine the potential for these objects to accelerate high-energy protons and power relativistic jets. Furthermore, we explore the variation in energy for accelerated protons from Sgr A* post-beta decay as a function of the quantum parameter $\epsilon$ and the magnetic field strength $B$, highlighting the observable signatures of LQG in astrophysical environments.

The motivation for this research lies in the possibility that LQG may significantly alter our understanding of highly energetic phenomena (such as AGNs, relativistic jet particle acceleration, etc.) in the vicinity of a BH. If quantum gravity effects impact the Penrose process, we may gain new insights into the observable properties of BH, including energy extraction, particle acceleration, and the role of magnetic fields in these processes. The exploration of Loop Quantum BHs (LQBHs) and their energy extraction mechanisms opens up a new avenue for theoretical and observational studies in BH physics, with potential implications for future experiments in observational astrophysics.

This study is also highly relevant in the context of current and future advancements in modern astrophysics. As gravitational wave astronomy \cite{Tarrant:2019nmn,Calcagni:2020tvw,Majumdar:2023bne} and Event Horizon Telescope (EHT) \cite{Islam:2022wck,Afrin:2022ztr,Kumar:2023jgh,Zhao:2024elr} observations continue to provide unprecedented views of BHs and their surroundings, the role of quantum effects in BH dynamics may become increasingly important. The growing interest in the study of BH mergers, the nature of BH horizons, and the behavior of accretion disks and relativistic jets necessitates an understanding that goes beyond classical general relativity. Quantum gravity models, like LQG, offer promising pathways to better understand phenomena that were once thought to be the domain of classical physics alone. The results of this study could inform the interpretation of future observations, such as the detection of gravitational waves from BH mergers or the imaging of BH shadows, and may provide key insights into the fundamental nature of spacetime itself.

This paper is organized as follows. Section \ref{Sec:Magnetized} introduces the metric of the LQBH and characterizes the surrounding magnetic field, along with a detailed examination of the ergosphere region and the event horizon limit, both expressed as functions of the parameter \(\epsilon\) (the product of the Immirzi parameter \(\gamma\) and the polymeric parameter \(\delta\)). Section \ref{Sec:Motion} focuses on the dynamics of charged particles in the LQBH background, setting the stage for an analysis of energy extraction. In \cref{Sec:Penrose}, we delve into the energy extraction mechanisms through the MPP, highlighting the impact of loop quantum gravity corrections on its efficiency. Section \ref{Sec:Apl} explores astrophysical applications, particularly focusing on constraining the mass and magnetic field of various BH candidates (such as SgrA*, M87*, NGC1052, and BZ) to serve as sources of high-energy protons. This section also examines the energy variation of accelerated protons after beta decay as a function of \(\epsilon\) for the magnetic field strengths \(B=100\;G\). Finally, \cref{Sec:conclusion} summarizes the work and presents concluding remarks. Throughout this paper, we use a spacetime metric signature \((- , + , + , +)\) and adopt geometric units with \(G = c = 1\).


\begin{figure*}[ht]
\minipage{0.54\textwidth}
\includegraphics[scale=0.87]{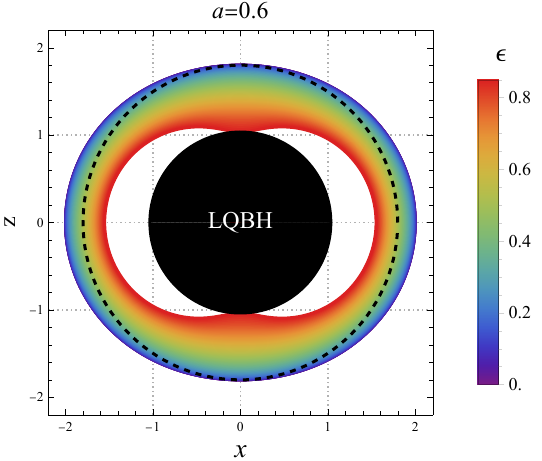} 
\endminipage
\minipage{0.46\textwidth}
\includegraphics[scale=0.8]{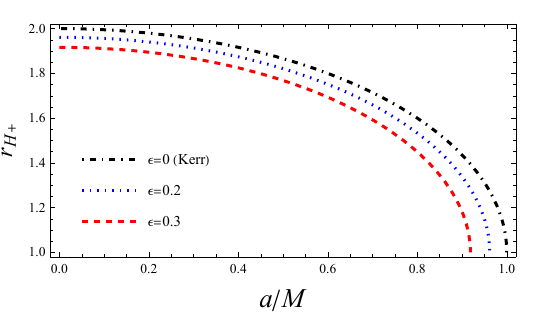}\vspace{-0.65cm}\\ %
\includegraphics[scale=0.8]{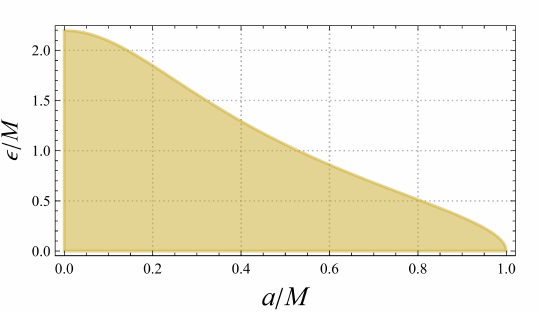}  
\endminipage
 \caption{\label{fig:erg-hor}    
\textit{Left panel:} Variation of the outer ergo-sphere $r_{s+}$ with $\theta$ for different values of the parameter $\epsilon$. The dotted line represents the event horizon $r_{H+}$ for $\epsilon = 0$, while the inner black disk corresponds to the event horizon in the extremal case.  
\textit{Right top panel:} Dependence of $r_{H+}$ on the spin parameter $a/M$ for various values of $\epsilon$.  
\textit{Right bottom panel:} Parameter space between the deformation parameter $\epsilon$ and the spin parameter $a/M$. The shaded region represents the theoretically allowed values of parameters $\epsilon$ and $a$ for which LQBHs exist.
}   
\end{figure*}

\section{Loop Quantum Black Hole Spacetime and its electromagnetic field }\label{Sec:Magnetized}

In this section, we examine the key properties of an intriguing model of rotating LQBHs \cite{Konoplya_2016,Ni_2016} immersed in a uniform external magnetic field ($B$), including the event horizon, static-limit-surface, and ergosphere. Notably, this BH is geodesically complete and free from curvature singularities at its core. The metric for this LQBH is presented in Boyer-Lindquist coordinates as follows \cite{Liu:2020ola}:
\begin{eqnarray}
ds^2 &=& -\frac{\Delta}{\Sigma}\left(dt - a \sin^2 \theta d\phi \right)^2 + \frac{\Sigma}{\Delta} dr^2 + \Sigma d\theta^2   
+ \frac{\sin^2 \theta}{\Sigma} \left( a dt - \left(k^2 + a^2 \right) d\phi \right)^2,
\end{eqnarray}
with  
\begin{eqnarray}
&\Delta = \frac{(r - r_+)(r - r_-)(r^2)}{(r + r_*)^2} + a^2 \ ,
\nonumber \\
&\Sigma = k^2(r) + a^2 \cos^2 \theta \ , 
\nonumber \\
&k^2 =\frac{r^4 + a_0^2}{(r + r_*)^2} \ ,
\end{eqnarray}
where $a$ is the spin parameter of the BH. The two horizons of the non-rotating LQBH are given by $r_{+} = 2M/(1 + P)^2$ and $r_{-} = 2M P^2/(1 + P)^2$. Additionally, $r_{*} = \sqrt{r_+ r_-} = 2 M P/(1 + P)^2$, where $M$ denotes the ADM (Arnowitt-Deser-Misner) mass of the solution. $P$ is the polymeric function and can be expressed as follows:

\begin{equation}
P = \frac{\sqrt{1 + \epsilon^2} - 1}{\sqrt{1 + \epsilon^2} + 1},
\end{equation}
where $\epsilon$ is the product of the Immirzi parameter $\gamma$ and the polymeric parameter $\delta$, and it must satisfy the condition $ \epsilon = \gamma \delta \ll 1$.

\begin{equation}
a_0 = \frac{A_{\text{min}}}{8 \pi},
\end{equation}
where $A_{\text{min}}$ corresponds to the minimum area gap in LQG. In addition, $A_{\text{min}}$ is connected to the Planck length $l_{\text{P}}$ by the expression

\begin{equation}
    A_{\text{min}} \simeq 4\sqrt{3} \pi \gamma  l_{\text{P}}^2 \ .
\end{equation}
As a result, $a_0$ scales with the Planck length $l_{\text{P}}$ and is anticipated to be extremely small. Thus, the influence of $a_0$ on spacetime should be negligible at observable scales. We will therefore set $a_0 = 0$ in this research.

Let's explore how the parameter $\epsilon$ affects the horizon and ergo-surface in the context of LQBH. Horizons and static-limit-surface are defined by the conditions $g^{rr} = 0$ and $g_{tt} = 0$, respectively. Based on these conditions, we can write the equations for the horizons and ergo-surfaces as follows,
\begin{equation} \label{eq:horizon}
    r_{H\pm}=\frac{1}{4}(r_++r_-)+r_{H1} \pm r_{H2}\ ,
\end{equation}
\begin{equation} \label{eq:ergosphere}
    r_{s\pm}=\frac{1}{4}(r_++r_-)+r_{s1}\pm r_{s2} \ ,
\end{equation}
where $r_{H1}$, $r_{H2}$, $r_{s1}$, and $r_{s2}$ are given by [\ref{A1}, \ref{A2}, \ref{A3}, \ref{A4}], respectively. It should be noted that when $\epsilon = 0$, Eq.~\eqref{eq:horizon} reduces to the horizon of the Kerr solution $r_{H\pm}=M\pm \sqrt{M^2-a^2}$. Fig.~\ref{fig:erg-hor} shows the effect of $\epsilon$ on the outer horizon (\textit{top right}) and outer static-limit-surface (\textit{left panel}). As $\epsilon$ increases, both the outer horizon and outer static-limit-surface shrink. However, it is important to note that for a fixed value of the spin parameter $a$, the overall ergosphere region (i.e. the space between the outer event horizon and outer static-limit-surface) increases as epsilon increases (see \cref{fig:erg-hor} left panel). 
The bottom right panel of \cref{fig:erg-hor} presents the permissible (shaded) values of parameters $\epsilon$ and $a$ for which LQBH exits.

In astrophysics, studying the effect of magnetic fields on the properties of BHs is crucial. Previous studies have shown that the magnetic field strongly influences the properties of a BH's accretion disk \cite{Blandford1977,PhysRevD.17.1518}. In this study, we assume the BH is surrounded by a weak magnetic field. Theoretical studies and observational data indicate that magnetic field strengths vary across different BHs; in some, the field is around $10^4$, while in others, it can reach up to $10^8$ (see, e.g., in Refs.~\cite{Piotrovich10,Baczko16,Daly:APJ:2019}). For example, the magnetic field strength in NGC1052 is in the range of $B\sim 200 - 8 \times 10^4~\rm{G}$, while in M87, it ranges from $1~\rm{G}$ to $30~\rm{G}$ (see, e.g., in Refs.~\cite{Baczko16,MF:2021ApJ,Narayan2021ApJ}). Although the magnetic field is weak, it has a strong effect on the motion of charged particles around the BH. Therefore, the motion of particles in the magnetic field surrounding a BH has been extensively studied, and various models have been developed to address different scenarios ~\cite[see, e.g.][]{Aliev02,Frolov10,Shaymatov20egb,Tursunov16,Shaymatov22a,Shaymatov21pdu,Shaymatov21c,Hussain17,Shaymatov22c,Shaymatov23GRG}.
For simplicity, we assume the magnetic field is uniform and aligned with the BH’s symmetry axis \cite{Wald74,Tursunov16,Shaymatov18a}. So, we can write the expression for the electromagnetic four-potential as follows: 
\begin{equation} \label{four-pot}
A^{\alpha}=C_1 \xi^{\alpha}_{(t)}+ C_2 \xi^{\alpha}_{(\varphi)}\, ,
\end{equation}
where $\xi^{\alpha}_{(t)}=(\partial/\partial t)^{\alpha}$ and
axial $\xi^{\alpha}_{(\varphi)}=(\partial/\partial \phi)^{\alpha}$ are killing vectors, and $C_1$ and $C_2$ are integration constants that define the property of field. Based on the characteristics of an asymptotically uniform magnetic field, we can define integration constants as $C_1 = aB$ and $C_2 = B/2$. 
Using Eq.~\eqref{four-pot}, we can define the components of the electromagnetic four-potentials as follows
\begin{align} \label{Eq:four-pot-at}
A_t &= -\frac{aB}{2\Sigma}\Big(2\Delta+\big[k(r)^2-a^2-\Delta\big]\sin^2\theta\Big) \, , \\
A_{\phi} &=-\frac{B \sin^2\theta}{2\Sigma}\Big(a^4-k(r)^4-2a^2\Delta+a^2\sin^2\theta\Delta\Big) \, .
\end{align}
In the following section, we focus on analyzing the motion of charged particles near the LQBH, which is placed in a uniform external magnetic field.\\

\section{Motion of Charged Particle in the Vicinity of LQBH} \label{Sec:Motion} 

Here, we investigate a charged particle motion around the LQBH using the Hamiltonian, which is defined by (see, for example~\cite{Misner73})
\begin{eqnarray}
H=\frac{1}{2}g^{\alpha\beta} \left(P_{\alpha}-qA_{\alpha}\right)\left(P_{\beta}-qA_{\beta}\right)\, ,
\end{eqnarray}
where $P_{\alpha}$ is the canonical momentum of a charged particle and $A_{\alpha}$ the four-vector potential of electromagnetic field. One can then connect the charged particle's four-momentum and its canonical momentum via the following equation
\begin{eqnarray}
p^{\alpha}=g^{\alpha\beta}\left(P_{\beta}-qA_{\beta}\right)\, . 
\end{eqnarray}
We can express the equation of motion for the charged particle using the Hamiltonian as follows:
\begin{eqnarray} 
\label{Eq:eqh1}
  \frac{dx^\alpha}{d\lambda} = \frac{\partial H}{\partial P_\alpha}\,   \mbox{~~and~~}
  \frac{dP_\alpha}{d\lambda} = - \frac{\partial H}{\partial x^\alpha}\, , 
\end{eqnarray}
where $(\lambda = \tau / m)$ denotes the affine parameter associated with $(\tau)$ describing the proper time of the particle. From the Eq.~(\ref{Eq:eqh1}), the constants of motion can be written as shown below:
\begin{eqnarray}
\label{Eq:en} P_t-qA_{t}&=&
g_{tt}p^{t} + g_{t\phi}p^{\phi}\, ,\\
 \label{Eq:ln}
P_{\phi}-qA_{\phi}&=& g_{\phi t}p^{t} +
g_{\phi\phi}p^{\phi}\, ,
\end{eqnarray}
where $P_t$ and $P_\phi$ correspond to the energy and angular momentum of a charged test particle, respectively, such that $P_t = -E$ and $P_\phi = L$.
Using two constants of motion given in Eqs.~(\ref{Eq:en},\ref{Eq:ln}) and the normalization condition $g_{\alpha\beta}p^{\alpha}p^{\beta}=-m^2$, the equation governing the timelike radial motion of a charged particle in the equatorial plane ($\theta = \pi/2$) can be expressed in terms of the effective potential in general form (see, e.g., \cite{Dadhich22b,Dadhich22a})
    \begin{eqnarray} \label{Eq:Veff}
        V_\text{eff} &=& \frac{a\, \beta \big(2 r^2-r (r_-+r_+)+r_- r_+\big)}{2 r^2} + \frac{a \big(r (r_- + r_+)-r_- r_+ \big) (r + rs)^2}{r^6 + 
 a^2 \big(r^2 - r_- r_+ + r (r_- + r_+)\big) (r + r_\star)^2}\cdot K\nonumber \\
&& +\sqrt{\frac{(r-r_-) (r-r_+)}{r^2}-\frac{a^2 (r+r_\star)^2 \big(r_- r_+-r (r_-+r_+)\big)^2}{a^2 r^2 (r+r_\star)^2 \big(r^2+r (r_-+r_+)-r_- r_+\big)+r^8}\   }  \nonumber \\
&& \times\sqrt{1+K^2 \Big( \frac{a^2 \big(r^2+r (r_-+r_+)-r_- r_+\big)}{r^2}+\frac{r^4}{(r+r_\star)^2} \Big)^{-1}}\, ,
    \end{eqnarray}
where $K$ is 
\begin{equation*}
    K =\mathcal{L}- \beta\cdot  \left(\frac{a^2 (r-r_-) (r-r_+)}{2r^2}+\frac{r^4}{2(r+r_\star)^2}\right)\, ,
\end{equation*}
with $\mathcal{L} = L/m$ and $\beta = q B G M/(m c^4)$ being the angular momentum per unit mass of the particle and a dimensionless magnetic field parameter. 

\begin{figure*}[ht]
\includegraphics[scale=0.48]{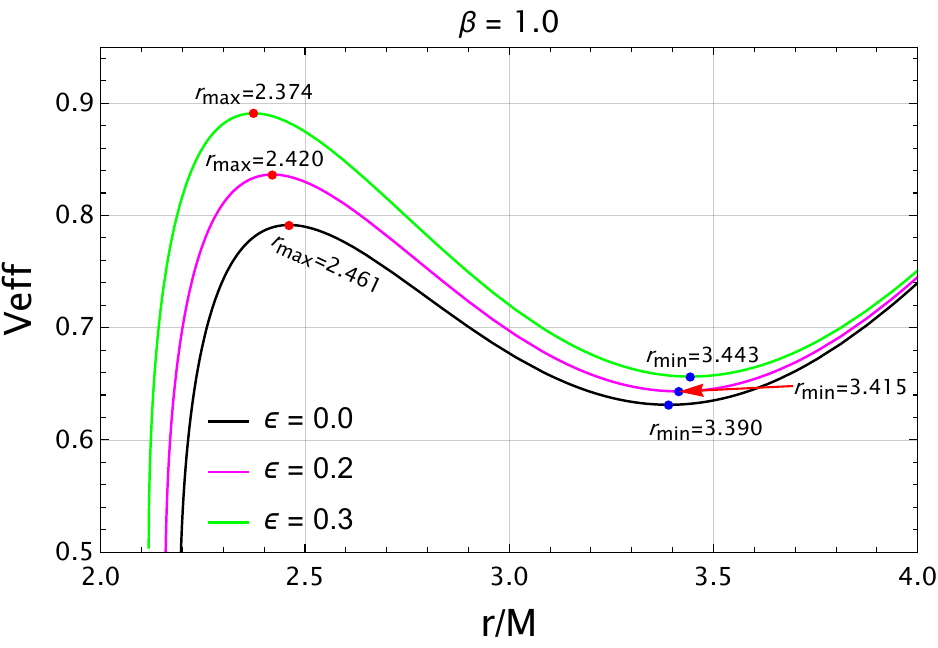}
\includegraphics[scale=0.48]{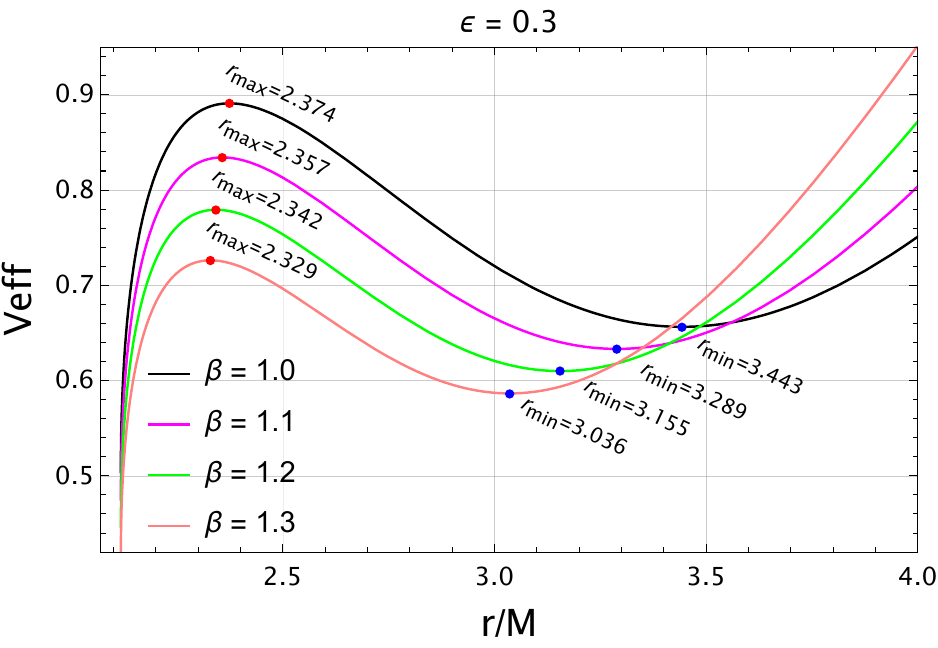}
	\caption{\label{fig:effpot} 
 The effective potential $V_{\text{eff}}$ is plotted as a function of $r/M$ for different values of $\epsilon$ (left panel) and the magnetic field parameter $\beta$ (right panel). In these plots, the spin parameter is set to $a = 0.8$.}
\end{figure*}

The effective potential $V_{eff}$ is a key tool for examining particle motion around a BH. The effective potential for a particle moving around the LQBH in the presence of an external magnetic field is shown in Fig.~\ref{fig:effpot}. The left and right panels of Fig.~\ref{fig:effpot} depict the radial variation of the effective potential for different values of $\epsilon$ and the magnetic field parameter $\beta$, where one parameter is fixed while the other varies. In the figure, $r_{\text{min}}$ and $r_{\text{max}}$ correspond to the stable and unstable circular orbits of the particle, respectively. The left panel of the figure depicts how variations in $\epsilon$ affect the effective potential of the system when $\beta$ is fixed. It is evident that as $\epsilon$ increases, the unstable orbit shifts to the left, i.e., to smaller values of $r/M$, while the stable orbit shifts to the right. In addition, increasing $\epsilon$ raises the height of the effective potential. This occurs because $\epsilon$ reduces the horizon radius $r_H$ (see Fig.~\ref{fig:erg-hor}). As $\epsilon$ increases, the horizon $r_H$ shrinks, enhancing the gravitational influence of the BH. This results in the formation of unstable orbits $r_{\text{max}}$ closer to the BH's center, as well as unstable orbits $r_{\text{min}}$ at greater distances.
The right panel depicts the effect of the magnetic field on the effective potential. As $\beta$ increases, both the stable orbit $r_{\text{min}}$ and the unstable orbit $r_{\text{max}}$ shift to the left, while the height of the effective potential decreases. This is a well-known effect of the magnetic field, indicating that a stronger magnetic field causes the particle to be more tightly confined towards the center of the BH.

Let us then analyze the motion of a charged particle on a stable circular orbit. In such motion, the radius vector $r$ remains constant, and the following condition must be satisfied for circular motion:
\begin{equation}
    V_{eff}(r)=\frac{\partial V_{eff}(r)}{\partial r}= 0 .
\end{equation}

For circular motion in the equatorial plane ($r = \text{const}$ and $\theta = \text{const}$), the four-velocity of the particle can be expressed as ${\bf u} \sim {\bf \xi}_{(t)} + \Omega {\bf \xi}_{(\phi)}$, where  
\begin{equation} \label{eq:four-vel}
    \Omega = \frac{d\phi}{dt} = \frac{u^{\phi}}{u^{t}}\, ,
\end{equation}
represents the angular velocity of the particle as measured by a distant observer at infinity. The angular velocity $\Omega$ of a timelike particle is bounded by the angular velocities of the inner and outer photon circular orbits, satisfying $\Omega_{-} < \Omega < \Omega_{+}$ (see details, e.g., in Refs.~\cite{1986ApJ...307...38P,Wagh1985ApJ}), where
\begin{eqnarray}
  \Omega_{\pm} =
  \frac{ \left(a(r_-+r_+)r-a r_\star^2\pm r^2 \sqrt{\Delta}\right) }{a^2  \left(r^2-r_\star^2+r (r_-+r_+)\right)+\frac{r^6}{(r+r_\star)^2}}\, .
\end{eqnarray}

\begin{figure}
\centering
\includegraphics[scale=0.55]{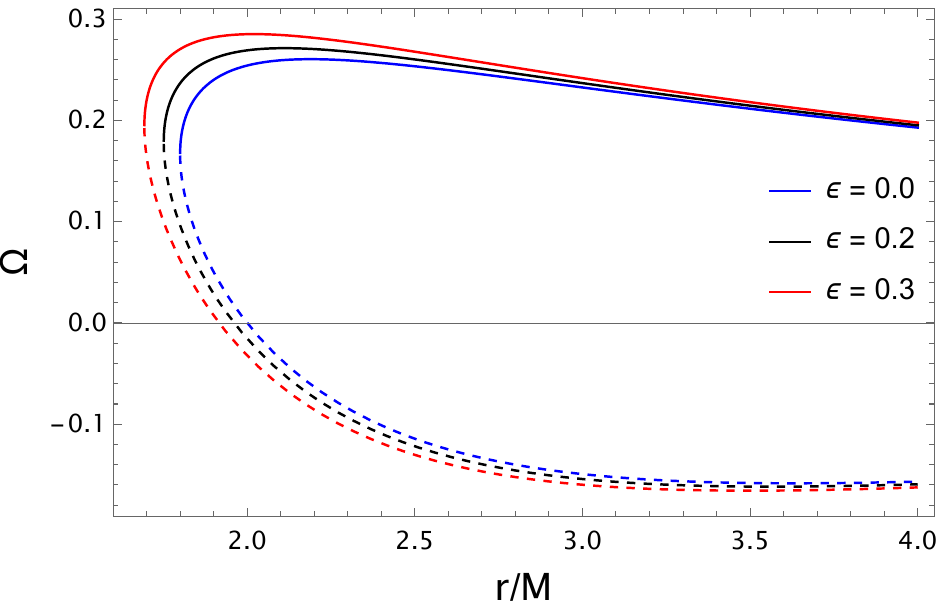}
	\caption{\label{fig:angular_vel} 
The angular velocity components $\Omega_+$ (solid lines) and $\Omega_-$ (dashed lines) are plotted against $r/M$ for different values of $\epsilon$, with the spin parameter fixed at $a = 0.6$. For simplicity, we set $\theta = \pi/2$.}
\end{figure}

We can examine the effect of $\epsilon$ on the angular velocity by referring to Fig.~\ref{fig:angular_vel}. The figure shows that $\Omega_+$ always remains positive, while $\Omega_-$ can take both positive and negative values. Moreover, it is clear that as $\epsilon$ increases, $\Omega_+$ increases, while $\Omega_-$ decreases. For the timelike circular orbit,  we can write the four momentum as 
$\pi_{\pm}=p^{t}(1,0,0,\Omega_{\pm})$.
The equation governing the circular motion of a timelike particle is derived by applying the normalization condition together with the expression for the particle's four momentum 
\begin{eqnarray}\label{Eq:W0}
\left(g_{\phi\phi}\pi_t^2+g_{t\phi}^2\right)\Omega^2&+&2g_{t\phi}\left(\pi_t^2+g_{tt}\right)\Omega
+g_{tt}\left(\pi_t^2+g_{tt}\right)=0\, ,
\end{eqnarray}
where we have denoted $\pi_t=-\left(\mathcal{E}+qA_{t}/m\right)$. Solving the above equation for $\Omega$ gives the angular velocity of the timelike particle (see details, for example ~\cite{1986ApJ...307...38P,Shaymatov:2022eyz})

\begin{eqnarray}\label{29}
\Omega=
\frac{ r^3 \sqrt{\pi_t^2 \big[\left(\pi_t^2-1\right) r^2+r (r_-+r_+)-r_\star^2\big] \Delta}+a [r (r_-+r_+)-r_\star^2] \big[\left(\pi_t^2-1\right) r^2+r (r_-+r_+)-r_\star^2\big]}{a^2 \Big[\pi_t^2 r^2 \left[r^2+r (r_-+r_+)-r_\star^2\right]+[r_\star^2-r (r_-+r_+)]^2\Big]+ \frac{\pi_t^2 r^8}{(r+r_\star)^2}}\, .\nonumber\\
\end{eqnarray}

\section{Harnessing Energy V\lowercase{ia} the Magnetic Penrose Process (MPP) from rotating LQBH}\label{Sec:Penrose}

In the ergoregion of a rotating BH, a particle can have negative energy, which allows for the possibility of extracting energy from the BH. The process was first theoretically described by Penrose \cite{Penrose:1969pc}, who proposed that a particle in the ergoregion could split into two parts. One part falls into the BH, while the other escapes. As a result, the energy of the escaping particle is greater than the energy of the original particle, effectively extracting energy from the BH. Further, we focus on the MPP \cite{Wagh1985ApJ,Wagh:1989zqa,Shaymatov24MPP1} to analyze the purely magnetic field effect on the energy extraction from LQBHs.
\begin{figure*}
\includegraphics[scale=0.5]{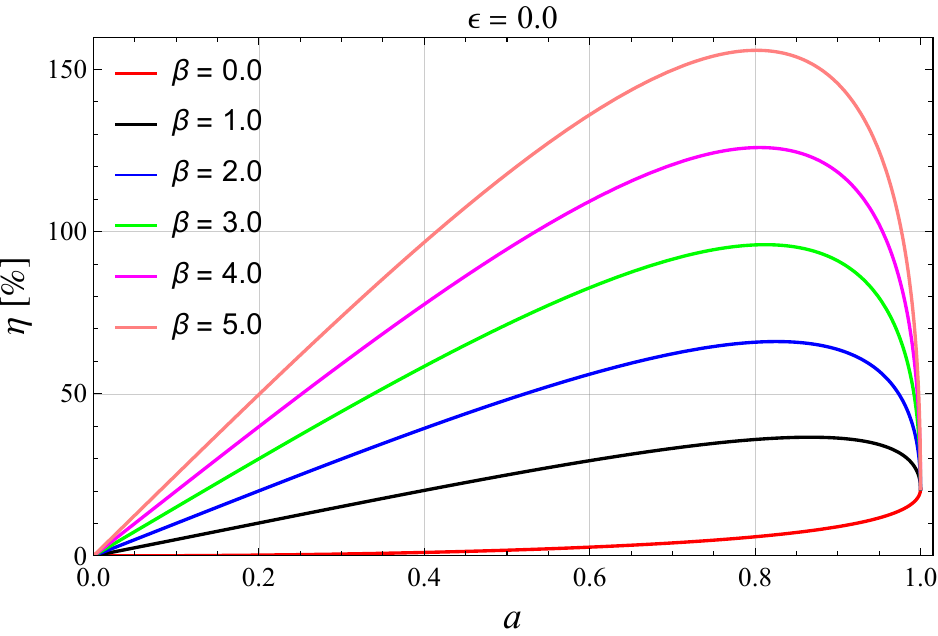}
\includegraphics[scale=0.5]{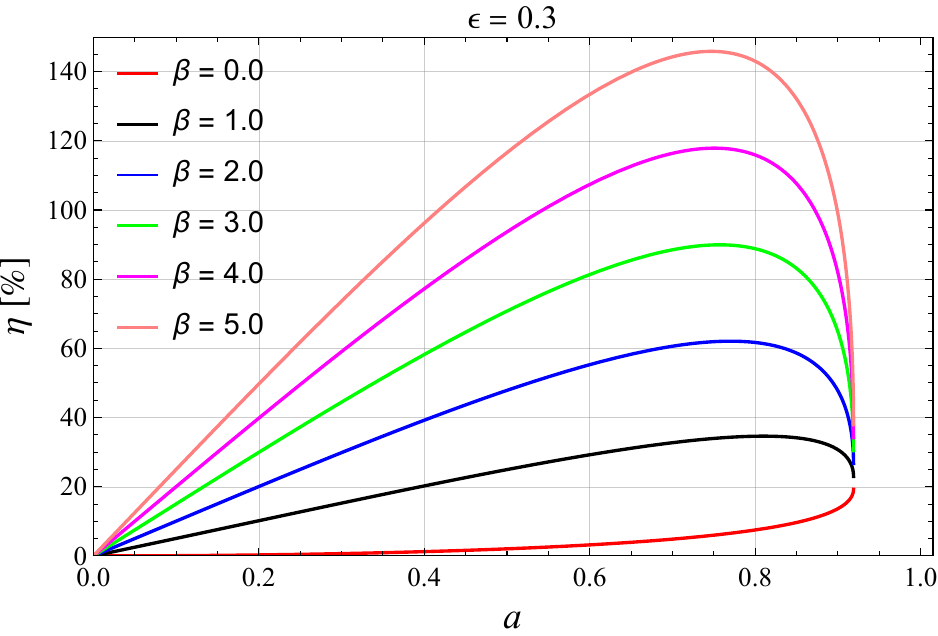}
\includegraphics[scale=0.5]{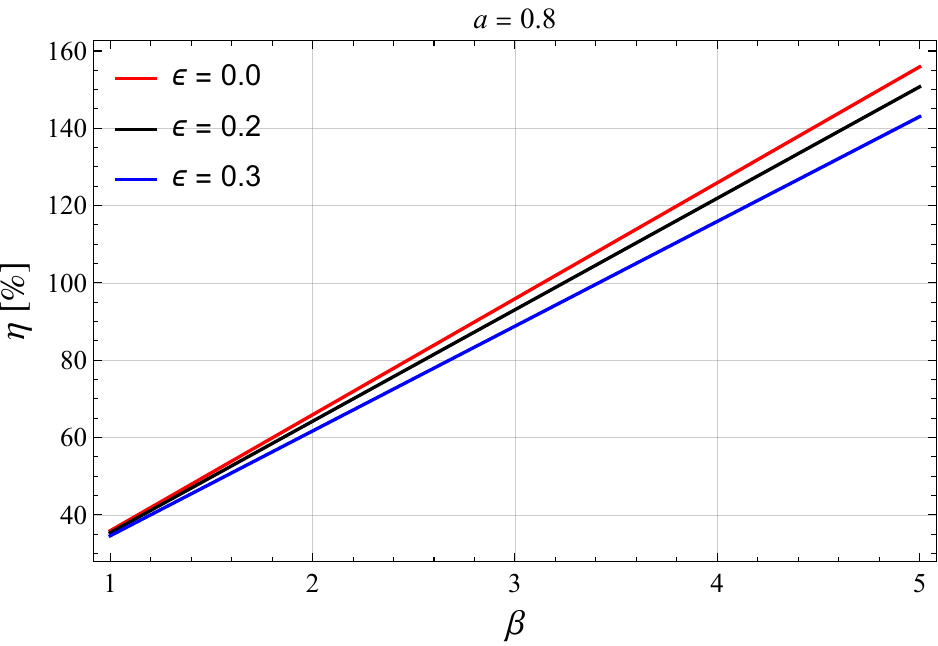}
\includegraphics[scale=0.5]{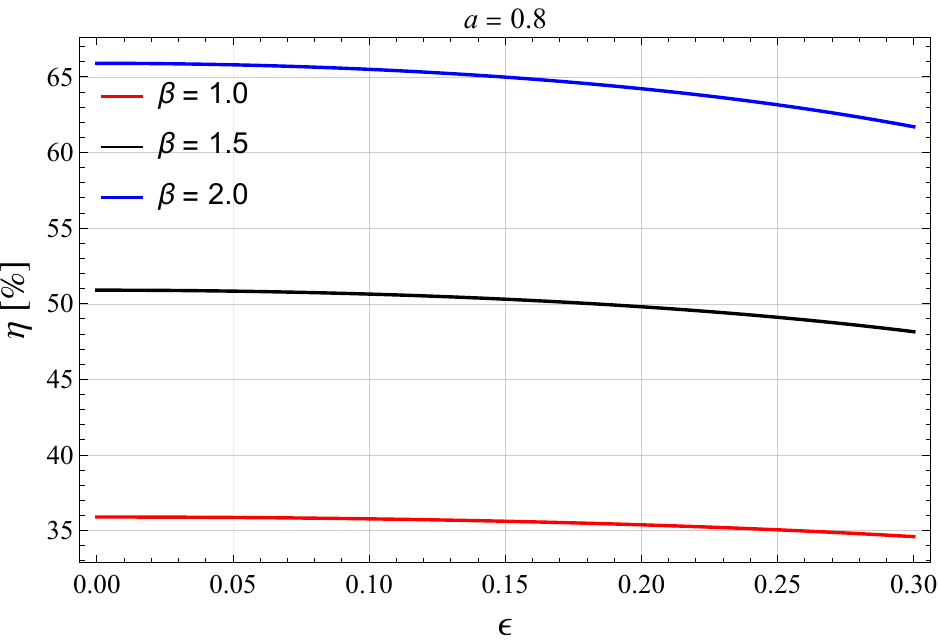}
\caption{\label{fig:en_eff1} The graphs show the impact of $\epsilon$ on the efficiency. The top row shows the relationship between the efficiency $\eta$ and the spin parameter $a$ for the two cases $\epsilon = 0.0$ and 
$\epsilon = 0.3$. In the bottom row, $\eta$ is plotted as a function of the magnetic field parameter $\beta$ for various possible values of $\epsilon$ (left), and as a function of $\epsilon$ for different values of $\beta$ (right) with fixed $a = 0.8$ . }
\end{figure*}

Consider a neutral particle $q_1 = 0$ that falls into a BH and splits into two particles within the ergoregion. Let us denote the energies and charges of the initial particle and the two resulting particles as $(E_1, q_1)$, $(E_2, q_2)$ and $ (E_3, q_3) $, respectively. We assume that the second particle, with mass $m_2$, falls into the BH carrying energy $E_2 < 0$, while the third particle, with mass $m_3$, escapes the BH with energy $E_3 = E_1 - E_2$, which exceeds the energy of the original particle, i.e., $E_3 > E_1$. The following conservation laws can be written for this process:
\begin{eqnarray} \label{Eq:con_laws}
    E_1&=&E_{2}+E_{3}\, ,\\
    L_1&=&L_{2}+L_{3}\, ,\\
    m_1&=&m_{2}+m_{3}\, ,\\
    q_1&=&q_2+q_3\, .
\end{eqnarray}
Following \cite{Blandford1977,Shaymatov:2022eyz},
the four-momentum can yield as \begin{eqnarray}\label{Eq:con_law}
m_1u_1^{\alpha}&=& m_2u_2^{\alpha}+m_3u_3^{\alpha}\, ,
\end{eqnarray}
satisfying the conservation laws. Using Eq.~\eqref{eq:four-vel}, we can express the relation $u_i^{\phi}=\Omega\, u_i^{t}=-\Omega \Lambda_i/\Gamma_i$  for four-velocity components of the particles. Consequently, Eq.~\eqref{Eq:con_law} can be rewritten in the following form
\begin{eqnarray}
\Omega_1m_1\Lambda_{1}\Gamma_2\Gamma_3=\Omega_2m_2\Lambda_{2}\Gamma_3\Gamma_1+\Omega_3m_3\Lambda_{3}\Gamma_2\Gamma_1\, ,
\end{eqnarray}
where  $\Lambda_i=\mathcal{E}_i +q_iA_{t}/m_i$ and $\Gamma_i=g_{tt}+g_{t\phi} \Omega_i $. After performing some simplifications, we arrive at the following equation:
\begin{eqnarray}
\frac{E_3+q_3A_{t}}{E_1+q_1A_{t}}=\left(\frac{\Omega_1\Gamma_2-\Omega_2\Gamma_1}{\Omega_3\Gamma_2-\Omega_2\Gamma_3}\right)\frac{\Gamma_3}{\Gamma_1}\, .
\end{eqnarray}
From the above equations, one can write the escaping particle's energy $E_3$ as follows
\begin{eqnarray}\label{Eq:E3E1}
E_3=\chi\left(E_1+q_1A_{t}\right)-q_3A_{t}\, ,
\end{eqnarray}
where we define $\chi$ as
\begin{eqnarray}\label{Eq:chi}
\chi=\left(\frac{\Omega_1-\Omega_2}{\Omega_3-\Omega_2}\right)\frac{\Gamma_3}{\Gamma_1}\, ,
\end{eqnarray}
with $\Omega_1= \Omega$,  $\Omega_2=\Omega_{-}$ and $\Omega_3=\Omega_{+}$. 
Now, we move on to determine the expression of efficiency:
\begin{eqnarray} \label{Eq:eta}
\eta=\frac{E_{gain}}{E_{initial}}=\frac{E_3-E_1}{E_1}\, .
\end{eqnarray}
Taking the initial condition $q_1=q_2+q_3=0$ and Eq.~\eqref{Eq:E3E1} into account, Eq.~\eqref{Eq:eta} can be written as follows:
\begin{eqnarray} \label{Eq:EnerEff}
\eta= \left(\frac{\Omega-\Omega_{-}}{\Omega_{+}-\Omega_{-}}\right)\left(\frac{g_{tt}+\Omega_{+}g_{t\phi}}{g_{tt}+\Omega\,g_{t\phi}}\right)-1-\frac{q_3A_t}{E_1}\, .
\end{eqnarray}
The closer the splitting process occurs to the horizon, the higher the efficiency. When the splitting happens at the horizon ($r_H$), the efficiency reaches its maximum value
\begin{eqnarray}\label{etamax}
    \eta_{max}&=&\frac{1}{2} \left(\sqrt{1-\frac{(r_H-r_-) (r_H-r_+)}{r_H^2}}-1\right) 
    +\beta \frac{a\,(2 r_H^2-r_H (r_-+r_+)+r_- r_+)}{2 r_H^2},\nonumber\\
\end{eqnarray}
where we have denoted 
\begin{equation}
    \beta = \frac{q_3 B G M}{c^2E_1}\sim \frac{q BGM}{m c^4}\, ,
\end{equation}
which, as stated earlier, represents the dimensionless magnetic field parameter. Here, we normalize the spin parameter as $a\to a/M$.

Let us now examine how the parameters $a$, $\beta$, and $\epsilon$ affect the efficiency. The top row of Fig.~\ref{fig:en_eff1} shows the efficiency $\eta$ as a function of $a$ for various values of $\beta$, under the conditions $\epsilon = 0.0$ and $\epsilon = 0.3$. The bottom row illustrates efficiency as a function of both $\epsilon$ and $\beta$, considering different values of these parameters. As shown in the top-right figure, when the spin parameter $a$ approaches its maximum value, $a \to a_{\text{max}}$ (approximately $a = 0.919$ in the top right), the efficiency is about 19.3\%. This is slightly lower than the 20.7\% efficiency observed in the Kerr case at $a = 1$ (as seen in the top-left figure). From the top row of the figure, it is evident that when $\epsilon = 0$, the maximum efficiency reaches approximately 150\%. In contrast, for $\epsilon = 0.3$, the efficiency decreases to around 140\%. Thus, an increase in $\epsilon$ leads to a reduction in both the maximum efficiency and the spin parameter. This trend is more apparent in the bottom row of Fig.~\ref{fig:en_eff1}. In the bottom-left figure, efficiency is plotted as a function of the magnetic parameter $\beta$ for various values of $\epsilon$. It is evident that as $\beta$ increases, efficiency also increases; however, as $\epsilon$ increases, the slope of the lines decreases. Similarly, in the bottom-right figure, efficiency is shown as a function of $\epsilon$ for different values of $\beta$. Here, we observe that efficiency increases with $\beta$ but decreases as $\epsilon$ becomes larger. The decrease in efficiency and spin parameters with increasing $\epsilon$ is attributed to the impact of $\epsilon$ on the ergosphere and horizon. This relationship, illustrating the reduction of horizon and ergosphere with increasing $\epsilon$, is depicted in Fig.~\ref{fig:erg-hor}.

\begin{figure*}
\includegraphics[scale=0.52]{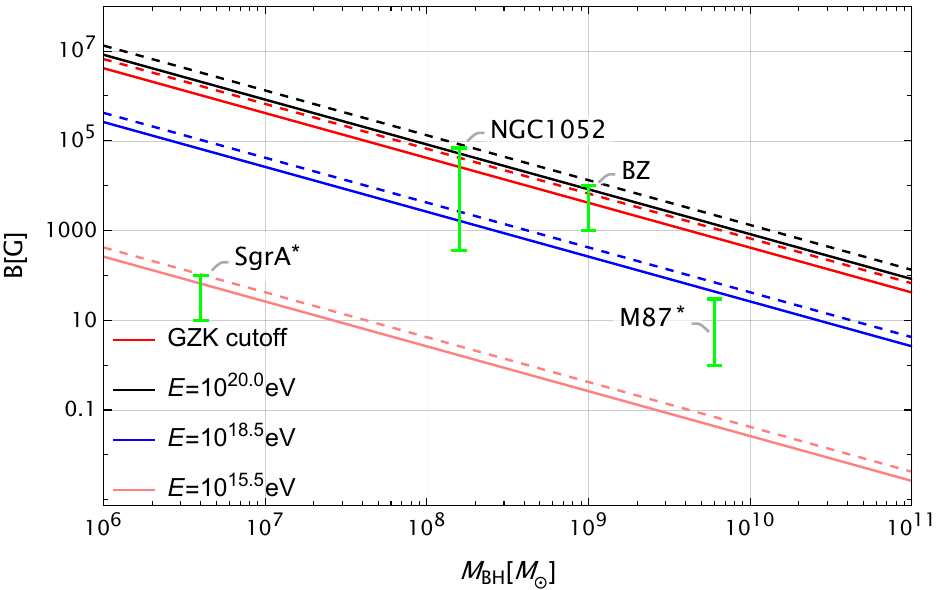}
\includegraphics[scale=0.50]{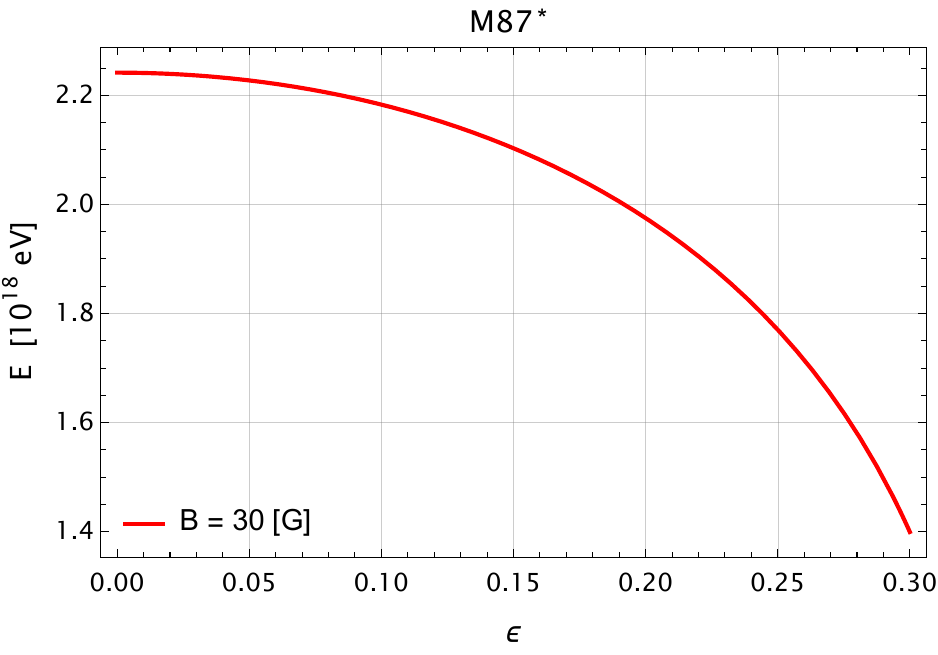}
\begin{center}
  \includegraphics[scale=0.5]{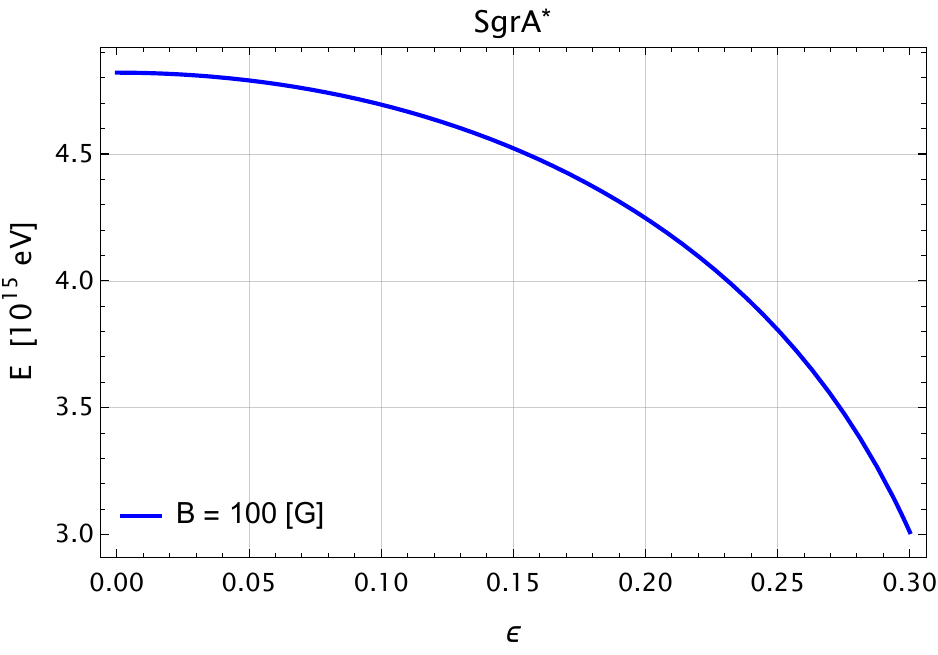}  
\end{center}
\caption{\label{fig:BMconstrain} The top left panel shows the constraint plot of BH mass $ M_{\text{BH}} $ and magnetic field $B$ for selected BH candidates (such as SgrA$^{\star}$, NGC1052, M87$^{\star}$, and 
 BZ), which could serve as sources of high-energy protons at different energy levels. The source labeled as BZ corresponds to a supermassive BH with a mass $10^9 M_\odot$ and a magnetic field strength ranging from $10^3$ G to $10^4$ G, consistent with the Blandford \& Znajek model of relativistic jets \cite{Blandford1977}. The lines depict various proton energies along with the GZK cutoff limit. The solid lines correspond to $\epsilon = 0$, representing a Kerr BH, while the dashed lines represent $\epsilon = 0.3$ case.
The top right and bottom panel display the energy $E$ of protons accelerated by M87$^{\star}$ and SgrA$^{\star}$ after beta decay as a function of $\epsilon$, plotted for the corresponding 
mass $M = 6.2 \times 10^9 \, M_\odot$ and $M = 4 \times 10^6 \, M_\odot$, respectively. In all plots, the spin parameter is set to $a = 0.9$.}
\end{figure*}

\section{Assessing the Efficiency of the MPP in LQBH\lowercase{s} Using M87*, S\MakeLowercase{gr}A*, NGC 1052, and BZ Observations \label{Sec:Apl}}

In this section, we explore the astrophysical applications of the MPP around LQBHs. 
Here, we focus on the amount of energy it can transfer to protons via MPP. This investigation offers valuable insights into the potential sources responsible for accelerating protons observed in cosmic rays. Suppose a neutron experiences beta decay very close to the surface of LQBH’s horizon. This process can be described as follows \cite{PierreAuger:2018qvk,Tursunov:2020juz,2022Symm...14..482T}:
\begin{eqnarray}
    n^0 \rightarrow p^+ + W^- \rightarrow p^+ + e^- + \overline{\nu}_e\, .
\end{eqnarray}
In this process, the initial particle is a neutron $n^0$, and the escaping particle is a proton $p^+$. Let's calculate the energy of the escaping proton. From Eqs.~\eqref{Eq:eta} and \eqref{etamax}, we can define the maximum energy of the escaping proton as follows

\begin{eqnarray}\label{EpMax}
    E_{max}^{p^+}&=&\frac{1}{2} \left(\sqrt{1-\frac{(r_H-r_-) (r_H-r_+)}{r_H^2}}+1\right)m_n c^2 
    +\frac{e BGM}{c^2}\cdot \frac{a\,(2 r_H^2-r_H (r_-+r_+)+r_- r_+)}{2 r_H^2},\nonumber
    \\
\end{eqnarray}
where $r_H$ is the horizon of the LQBH.

Let us now explore how $\epsilon$ influences the energy of the escaping particle. To illustrate this, Table \ref{EscapeEnergy} presents the calculated energy of the escaping proton for various values of epsilon across different BH candidates.
From the Table \ref{EscapeEnergy}, we observe that the acceleration capabilities of these BHs depend on their mass, the magnetic field strength in their vicinity, and the parameter $\epsilon$. As the mass and magnetic field strength increase, the energy of the accelerated particle also increases. However, an increase in $\epsilon$ reduces the particle's energy. 

\begin{table}[] \vspace{0.5cm}
\resizebox{\textwidth}{!}{
\begin{tabular}{|c|c|c|lllll|}
\hline
\multirow{2}{*}{Source} & \multirow{2}{*}{\begin{tabular}[c]{@{}c@{}}Mass\\ {[}$M_\odot${]}\end{tabular}} & \multirow{2}{*}{\begin{tabular}[c]{@{}c@{}}B\\ {[}G{]}\end{tabular}}      & \multicolumn{5}{c|}{$E_{p^+}$ {[}eV{]}}                                                                                            \\ \cline{4-8} 
                      &                            &                         & \multicolumn{1}{c|}{$\epsilon=0.0$}   & \multicolumn{1}{c|}{$\epsilon=0.1$} & \multicolumn{1}{c|}{$\epsilon=0.2$} & \multicolumn{1}{c|}{$\epsilon=0.3$} & \multicolumn{1}{c|}{ }  \\ \hline
 $\text{SgrA}^\star$                  & $4.0\times 10^6$   & 100                     & \multicolumn{1}{l|}{4.094} & \multicolumn{1}{l|}{4.088} & \multicolumn{1}{l|}{4.071} & \multicolumn{1}{l|}{4.041} & $\times 10^{15}$ \\ \hline
M87$^\star$                   & $6.2\times 10^9$  & 30                      & \multicolumn{1}{l|}{1.904} & \multicolumn{1}{l|}{1.901} & \multicolumn{1}{l|}{1.893} & \multicolumn{1}{l|}{1.879} & $\times 10^{18}$ \\ \hline
BZ                    & $1.0\times10^9$      & $10^4$   & \multicolumn{1}{l|}{1.024} & \multicolumn{1}{l|}{1.022} & \multicolumn{1}{l|}{1.018} & \multicolumn{1}{l|}{1.010} & $\times 10^{20}$ \\ \hline
NGC 1052              & $1.6\times 10^8$ & $8\times10^4$ & \multicolumn{1}{l|}{1.297} & \multicolumn{1}{l|}{1.296} & \multicolumn{1}{l|}{1.291} & \multicolumn{1}{l|}{1.281} & $\times 10^{20}$ \\ \hline
\end{tabular}
}
\caption{The table shows the impact of $\epsilon$ to the energy of escaping protons accelerated by different BHs. The energy of protons are calculated for various values of $\epsilon$. In this calculations, we set $a = 0.5$, which is the dimensionless spin parameter.}
 \label{EscapeEnergy}
\end{table}

For a more in-depth analysis of the contribution of BH candidates to the cosmic ray spectrum through the production of high-energy cosmic rays, the top left panel of Fig.~\ref{fig:BMconstrain}  presents the current constraints on their masses and magnetic field strengths, along with the critical energy points in the cosmic ray spectrum.
The vertical green lines in the figure represent the range of magnetic field strengths inferred from observational data for various BH candidates (see, e.g., \cite{MF:2021ApJ,Eckart_2012,Eatough_2013,Baczko16}). The solid black, blue and pink lines shows the energy of the escaping particle when $\epsilon = 0$, indicating the transition of the LQBH to a Kerr BH. 
The red lines indicate the Greisen-Zatsepin-Kuzmin (GZK) limit, commonly known as the cutoff effect. This limit defines the maximum energy that protons can achieve while traveling across the intergalactic medium from distant galaxies to our galaxy. Theoretically, this cutoff energy is estimated to be approximately $5 \times 10^{19} \, \text{eV}$ \cite{Greisen1966}. This limit arises from interactions between high-energy protons and the cosmic microwave background radiation over large-scale intergalactic distances.
The dashed lines illustrate the case of $\epsilon \neq 0$, showing the mass and magnetic field strength required for a BH to accelerate particles to these energies as an LQBH.
The figure demonstrates that as $\epsilon$ increases, a stronger magnetic field is required to accelerate a particle to the given energy. Additionally, it highlights that $\text{SgrA}^\star$ serves as a source of cosmic rays at the knee energy, while NGC 1052 is associated with cosmic rays at the ankle energy or in the ultra-high-energy range.

These energy lines signify critical features in the cosmic ray spectrum. For instance, beyond the knee energy ($10^{15} \text{--} 10^{16} \, \text{eV}$), the flux of cosmic ray particles drops significantly, indicating a sharp transition in the spectrum. In contrast, beyond the ankle energy ($10^{18.5} \, \text{eV}$), the spectrum flattens, reflecting a change in the dominant sources. The top right and bottom panel of Fig.~\ref{fig:BMconstrain} show the effect of $\epsilon$ on the energy of protons accelerated by $\text{SgrA}^\star$ and $\text{M87}^\star$. It can be seen from the figure that as $\epsilon$ increases, the energy of the protons decreases.





\section{Conclusions}
\label{Sec:conclusion}
In this study, we investigated the effects of LQG corrections on the MPP and its implications for particle motion and energy extraction around LQBHs. By placing an LQBH in an external asymptotically uniform magnetic field, we analyzed the impact of the quantum (correction) parameter $\epsilon$ on the efficiency of the process and the dynamics of particles in its vicinity. Furthermore, we explored the particle acceleration capabilities of the MPP by calculating the energy of accelerated protons for various BH candidates, considering current observational constraints on their magnetic field strengths and masses. We analyzed the impact of the parameter $\epsilon$ on the horizon and ergosphere of the LQBH, showing that both the outer event horizon and the outer static limit surface contract as $\epsilon$ increases for a fixed spin parameter $a$ (see Fig.~\ref{fig:erg-hor}). However, the ergosphere region expands under these conditions. A similar trend is observed for the ergosphere when $\epsilon$ is held constant, and $a$ increases.

To investigate the effect of $\epsilon$ on the dynamics of charged particle, we defined the effective potential $V_{eff}$. Our analysis revealed that as $\epsilon$ increases, unstable orbits contract, stable orbits expand, and the height of the effective potential increases. Conversely, under the influence of a magnetic field, the height of the effective potential decreases, and both stable and unstable orbits are compressed closer to the center (see Fig.~\ref{fig:effpot}).

From an astrophysical perspective, LQBHs are considered important objects for studying quantum effects in strong gravitational fields. Therefore, studying their energetic properties, such as particle acceleration, energy extraction, and the influence of magnetic fields on these processes, is crucial for understanding the nature and observable parameters of LQBHs. This research can provide a foundation for future studies on the observation and detection of LQBHs. Therefore, we utilized the MPP to explore the energetic properties of LQBHs. For the LQBH, the efficiency of the MPP can also exceed 100\%; however, we have demonstrated that increasing $\epsilon$ slightly reduces the efficiency. In the extreme spin case, we found that when $\epsilon = 0$, i.e., for a Kerr BH, the efficiency is the well-known 20.7\%, whereas, for $\epsilon = 0.3$, it decreases to 19.3\%. But, with the increasing of magnetic field, the efficiency always increases (see Fig.~\ref{fig:en_eff1}).

Additionally, we explored the acceleration properties of the MPP and evaluated the energy of escaping protons for various BHs (see Table \ref{EscapeEnergy}). To illustrate the role of various BHs in producing high-energy cosmic rays, we estimated their cosmic ray-producing capacity at key points in the cosmic ray spectrum, based on current constraints on their masses and magnetic fields (see, e.g.,\cite{MF:2021ApJ,Eckart_2012,Eatough_2013,Baczko16}). We demonstrated that these BHs, as LQBHs, would require a stronger magnetic field to accelerate particles to the energies achieved by Kerr BHs (see the left panel in Fig.~\ref{fig:BMconstrain}). Furthermore, a stronger magnetic field is also necessary to generate energy at the GZK cutoff limit under the influence of $\epsilon$. We analyzed the effect of $\epsilon$ on the capability of the BH $\text{SgrA}^\star$ at the center of our Milky Way Galaxy to accelerate protons and found that the energy of the accelerated protons decreases as $\epsilon$ increases (see the right panel in Fig.~\ref{fig:BMconstrain}).

From an astrophysical perspective, understanding what properties distinguish LQBHs from classical BHs is crucial. This requires examining the effects of quantum corrections on the BH horizon, ergosphere, and the energetic processes occurring in its vicinity. However, due to the vast distances to BHs, distinguishing between classical and quantum effects through direct observations remains a significant challenge.

Building on these theoretical foundations, the approach adopted in this study is critical for uncovering the unique characteristics of LQBHs. By analyzing the influence of the LQG correction parameter $\epsilon$ on the MPP, our work provides deeper insights into the distinctive energetic properties of LQBHs. This research not only enhances our understanding of quantum gravity effects in astrophysical observations but also lays the groundwork for future studies on the detectability and observational signatures of LQBHs. Given the importance of high-energy processes around astrophysical BHs, our findings are astrophysically significant because they do not preclude the influence of LQBHs, which could be sources of highly energetic phenomena and play a crucial role in explaining these powerful events.

%

\appendix
\section{The expressions of $r_{H1}$, $r_{H2}$, $r_{s1}$, and $r_{s2}$}

The generic forms of $r_{H1}$ and $r_{H2}$ are expressed as:
\begin{eqnarray}
r_{H1}&=&\frac{1}{2} \left[\frac{1}{3} \left(a^2+r_+ r_-\right)-a^2+\frac{1}{4} \left(r_++r_-\right)^2-r_+ r_-+\frac{A}{3 \sqrt[3]{2}}+B\right]^{\frac{1}{2}},  \label{A1} \\
r_{H2}&=&\frac{1}{2} \Bigg[\frac{-4 \left(r_++r_-\right) \left(a^2+r_+ r_-\right)-16 a^2 r_*+\left(r_++r_-\right)^3}{8 r_{H1}} 
-\frac{1}{3} \left(a^2+r_+ r_-\right)-a^2\nonumber\\
&&+\frac{1}{2} \left(r_++r_-\right)^2-r_+ r_--\frac{A}{3 \sqrt[3]{2}}-B\Bigg]^{\frac{1}{2}}, \label{A2}
\end{eqnarray}
where
\begin{eqnarray*}
A&=&\sqrt[3]{C+D},\\
B&=& \frac{\sqrt[3]{2}E}{3 A},\\
C&=&\sqrt{D^2 -4 E^3}, \\
D&=&108 a^4 r_*^2-72 a^2 r_*^2 \left(a^2+r_+ r_-\right)+18 a^2 \left(r_++r_-\right) r_* \left(a^2+r_+ r_-\right)+2 \left(a^2+r_+ r_-\right)^3\nonumber\\
&&+27 a^2 \left(r_++r_-\right)^2 r_*^2, \nonumber\\
E&=&12 a^2 r_*^2+6 a^2 \left(r_++r_-\right) r_*+\left(a^2+r_+ r_-\right)^2.
\end{eqnarray*}
The expressions for $r_{s1}$ and $r_{s2}$ are given by
\begin{eqnarray}
r_{s1}&=&\frac{1}{2} \left[-a^2 \cos ^2\theta+\frac{1}{3} \left(a^2 \cos ^2\theta+r_+ r_-\right)+\frac{1}{4} \left(r_++r_-\right)^2-r_+ r_-+\frac{\mathcal{C}}{3 \sqrt[3]{\mathcal{A}+\mathcal{B}}}+\frac{\sqrt[3]{\mathcal{A}+\mathcal{B}}}{3 \sqrt[3]{2}}\right]^{\frac{1}{2}}, \nonumber\\
\label{A3}\\
r_{s2}&=&\frac{1}{2} \left[\frac{1}{8 r_{s1}}\left(\left(r_++r_-\right)^3-4 \left(r_++r_-\right) \left(a^2 \cos ^2\theta+r_- r_+\right)-16 a^2 r_* \cos ^2\theta \right)-a^2 \cos ^2\theta\right. \nonumber\\
&&\left.-\frac{1}{3} \left(a^2 \cos ^2\theta+r_+ r_-\right)-r_+r_-
+\frac{1}{2} \left(r_++r_-\right)^2-\frac{\mathcal{C}}{3 \sqrt[3]{\mathcal{A}+\mathcal{B}}}-\frac{\sqrt[3]{\mathcal{A}+\mathcal{B}}}{3 \sqrt[3]{2}}\right]^{\frac{1}{2}}, \label{A4}
\end{eqnarray}
where
\begin{eqnarray}
\mathcal{A}=&&\sqrt{\mathcal{B}^2 -2\mathcal{C}^3},   \nonumber\\  \nonumber\\
\mathcal{B}=&&108 a^4 r_*^2 \cos ^4\theta +27 a^2 \left(r_++r_-\right)^2 r_*^2 \cos ^2\theta-72 a^2 r_*^2 \cos ^2\theta \left(a^2 \cos ^2\theta+r_+ r_-\right)\nonumber\\
&&+2 \left(a^2 \cos ^2\theta +r_+ r_-\right)^3
+18 a^2 \left(r_++r_-\right) r_* \cos ^2\theta \left(a^2 \cos ^2\theta+r_+ r_-\right),\nonumber\\  \nonumber\\
\mathcal{C}=&&\sqrt[3]{2} \left[12 a^2 r_*^2 \cos ^2\theta +6 a^2 \left(r_++r_-\right) r_* \cos ^2\theta +\left(a^2 \cos ^2\theta +r_+ r_-\right)^2\right]. \nonumber
\end{eqnarray}

\acknowledgments
This work is supported by the National Natural Science Foundation of China under Grant No. W2433018 and No. 12275238, the National Key Research and Development Program of China under Grant No. 2020YFC2201503, and the Zhejiang Provincial Natural Science Foundation of China under Grants No. LR21A050001 and No. LY20A050002. PS acknowledges the Vellore Institute of Technology for providing financial support through its Seed Grant (No. SG20230079), year 2023. 

%
\bibliographystyle{JHEP}
\bibliography{Ref.bib}

\end{document}